\documentclass[aps,prb,superscriptaddress,twocolumn,10pt]{revtex4-1}
\usepackage[T1]{fontenc}
\usepackage{graphicx}
\usepackage{lineno}

\begin{document}

\title{STM-induced desorption and lithographic patterning of Cl-Si(100)-(2$\times$1)}
\author{K.J. Dwyer}
\email{kdwyer2@lps.umd.edu}
\affiliation{Department of Physics, University of Maryland, College Park, MD 20740, USA.}
\author{Michael Dreyer}
\affiliation{Department of Physics, University of Maryland, College Park, MD 20740, USA.}
\author{R.E. Butera}
\email{rbutera@lps.umd.edu}
\affiliation{Laboratory for Physical Sciences, 8050 Greenmead Drive, College Park, MD 20740, USA.}



\begin{abstract}
We investigated STM-induced chlorine desorption and lithographic patterning of Cl-terminated Si(100)-(2$\times$1) surfaces at sample temperatures from \mbox{4 K} to \mbox{600 K}. STM lithography has previously focused on hydrogen-based chemistry for donor device fabrication. Here, to develop halogen-based chemistries for fabricating acceptor-based devices, we substituted the hydrogen resist with chlorine. Lithographic patterning was explored using both field emission patterning to desorb chlorine from large areas as well as atomic precision patterning to desorb chlorine along one to two dimer rows at a time. We varied the experimental parameters for lithographic patterning and found a positive correlation between pattern line widths and both sample bias voltage and total electron dose. Finally, the use of chlorine, bromine, and iodine as lithographic resists to broaden the range of available chemistries for future device fabrication utilizing halogen-based dopant precursors is discussed.
\end{abstract}

\maketitle


\section{Introduction}
Hydrogen depassivation lithography is a well-established technique for the fabrication of atomic-scale electronic devices in silicon\cite{Walsh:2009, Zyvex_HDL:2014}. An atomic layer of H passivating a Si(100) surface (H-Si(100)) is utilized as a resist for lithographic patterning. The H atoms can be selectively desorbed using the tip of a scanning tunneling microscope (STM) in an ultra-high vacuum (UHV) environment\cite{Shen1590}. Highly reactive Si dangling bonds exposed in the patterned area may then react with an appropriately chosen precursor molecule such as PH$_{3}$. This enables fabrication techniques including selective-area chemical vapor deposition\cite{Abeln_SACVD:1998} and patterned, atomic-layer epitaxy\cite{Zyvex_PALE:2011}. Through careful control of the precursor reaction kinetics\cite{Warschkow_PH3:2016}, placement of single phosphorus atoms on Si(100) with near atomic precision has been demonstrated for use as single atom transistors\cite{Fuechsle_Atom:2012} and donor-based qubits\cite{Broome-2Q:2018} in Si.

Recently, several proposals were put forth for the fabrication of boron and aluminum dopant-based devices in Si ranging from acceptor-based qubits\cite{Salfi:2016, Salfi_PRL:2016} to superconducting Si devices\cite{Shim:2014, Blase:2009}. To our knowledge, no H-based acceptor precursor analogous to PH$_{3}$ has been identified that is also compatible with STM-based lithography and atomic-precision device fabrication. However, BCl$_{3}$ has been used to heavily dope Si with boron by means of gas immersion laser doping (GILD) to the point of superconductivity\cite{CAMMILLERI200875}. Additionally, several reports have demonstrated room temperature adsorption of BCl$_{3}$ and AlCl$_{3}$ on Si(100)\cite{Ferguson:2009} and Ge(100)\cite{Youn:2008}, pointing to the viability of halogen-based acceptor precursors for use in atomic-precision fabrication. Halogen-based precursors likely necessitate the use of a halogen-based resist to prevent halogen reactions with a lithographically patterned H resist, thus destroying the pattern\cite{Koleske_JAP:1992, Gates_JPC:1992, Mendicino01061993}. Moreover, adsorbed Cl atoms were recently shown to effectively inhibit further adsorption of BCl$_{3}$ during atomic layer deposition experiments\cite{Pilli:2018}, thereby confirming its potential for use as an effective mask for halogenated acceptor precursor compounds.

In this paper we investigated STM-induced desorption and lithographic patterning of Cl-Si(100)-(2$\times$1) from \mbox{4 K} to \mbox{600 K}. We observed tip-induced reactions at both positive and negative sample biases, $V_{B}$, corresponding to electron and hole injection, respectively. Positive $V_{B}$ values above a threshold voltage resulted in a field emission patterning mode useful for desorbing large areas of Cl, while negative $V_{B}$ values produced near atomic-precision patterning. We found the patterning process to be more facile at elevated temperatures. Patterned areas remained stable up to \mbox{600 K}, but we observed both inter- and intradimer diffusion of remnant adsorbed Cl atoms within the patterned areas. Finally, the experimental parameters for lithographic patterning were varied revealing a positive linear dependence of the pattern feature line widths on $V_{B}$ and the total electron dose, $d_{e}$.

\begin{figure}[b!]
\centering
\includegraphics[width=0.33\textwidth]{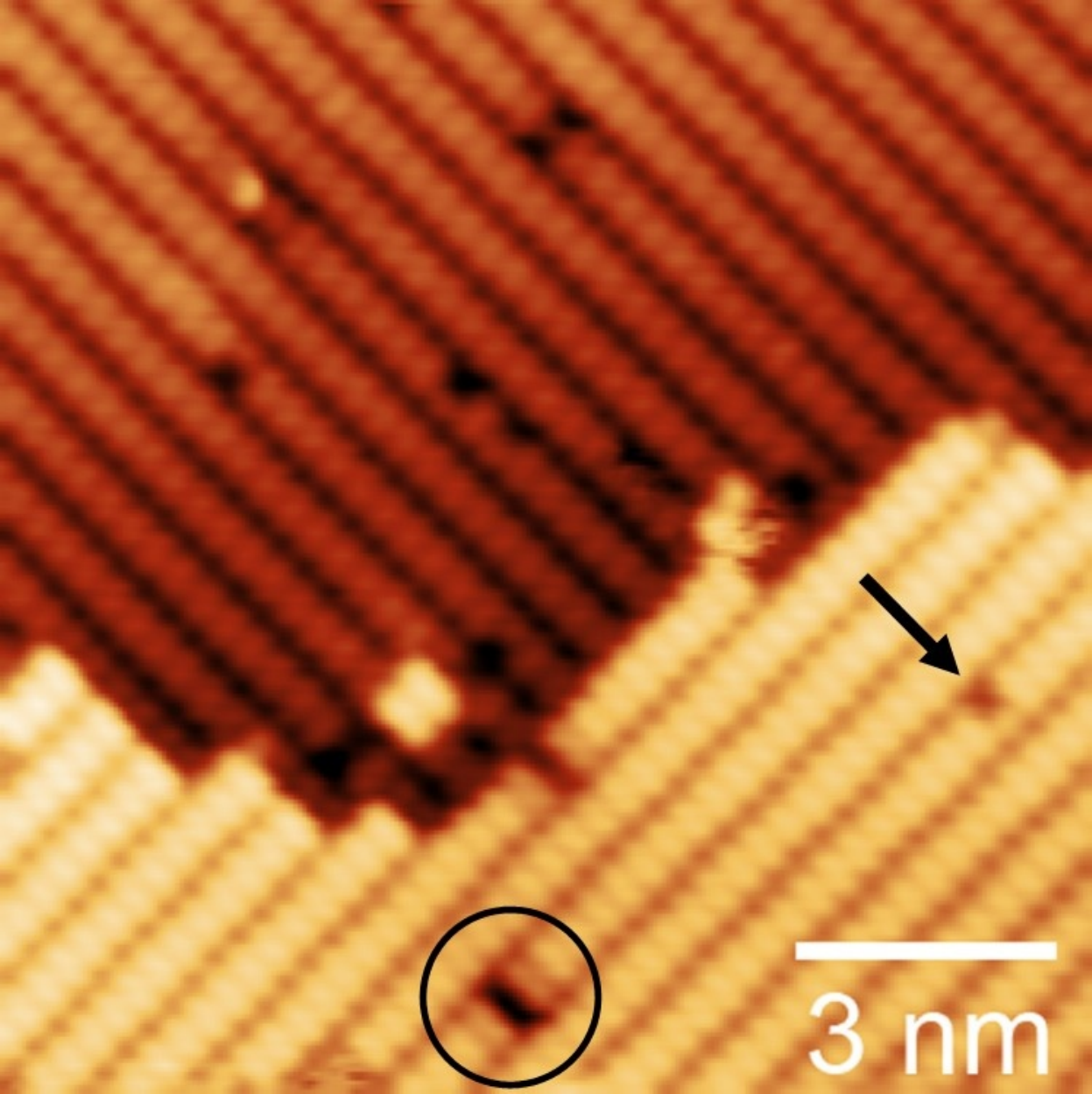}\hfill
\caption{Filled-state STM image ($-$1.7 V, 0.8 nA) of a fully terminated Cl-Si(100)-(2$\times$1) surface at \mbox{400 K}. The (2$\times$1) reconstruction of the bare surface is maintained with one Cl atom passivating each Si dangling bond. A single atomic step runs across the center of the image. The arrow identifies a single Cl vacancy, while the circle identifies a Si dimer vacancy.}
\label{cl_si}
\end{figure}

\section{Results and Discussion}

STM-induced desorption of adsorbed Cl atoms from a Si(100) surface was briefly mentioned within a study exploring the adsorption of TiCl$_{4}$ on STM patterned H-Si(100)\cite{Mitsui:1999}. Outside of this brief mention, no further investigations of STM-induced desorption or lithographic patterning of halogen-terminated Si(100) have been reported to the best of our knowledge. With recent theoretical work exploring the use of Cl-Si(100) as a patternable resist to foster truly atomic-precision placement of phosphorous atoms in silicon\cite{Pavlova:2018} and a desire to develop halogen-based chemistry for acceptor doping through STM fabrication, a more thorough investigation of the STM lithography process on Cl-Si(100) was warranted.

The results discussed here represent experiments from two UHV STM systems using Si(100) substrates of several different doping densities. One STM system was used to perform experiments at room (approximately \mbox{293 K}) and elevated temperatures with either higher doped ($\rho =$ \mbox{0.001 $\Omega\cdot$cm} to $\rho =$ \mbox{0.002 $\Omega\cdot$cm}) or lower doped ($\rho =$ \mbox{1 $\Omega\cdot$cm} to $\rho =$ \mbox{10 $\Omega\cdot$cm}) substrates. A separate cryogenic STM system was used to perform low-temperature experiments using high purity Si epilayer substrates.

A fully terminated Cl-Si(100)-(2$\times$1) surface can be seen in the representative filled-state STM image shown in Fig.~\ref{cl_si}. The image was acquired with the sample at \mbox{400 K} after a Cl saturation exposure on a lower doped substrate. Cl$_{2}$ dissociatively chemisorbs on Si(100) and each adsorbed Cl atom terminates the dangling bond of a single Si atom, maintaining the (2$\times$1) reconstruction of the bare surface\cite{Lyub:1998}. A single atomic-height step runs through the middle of the image and several common defects are identified including single Cl vacancies (arrow) and a single Si dimer vacancy (circle). With the exception of the Cl vacancies, each Si dimer is fully Cl-terminated.

\begin{figure}[t!]
\centering
\includegraphics[width=\columnwidth]{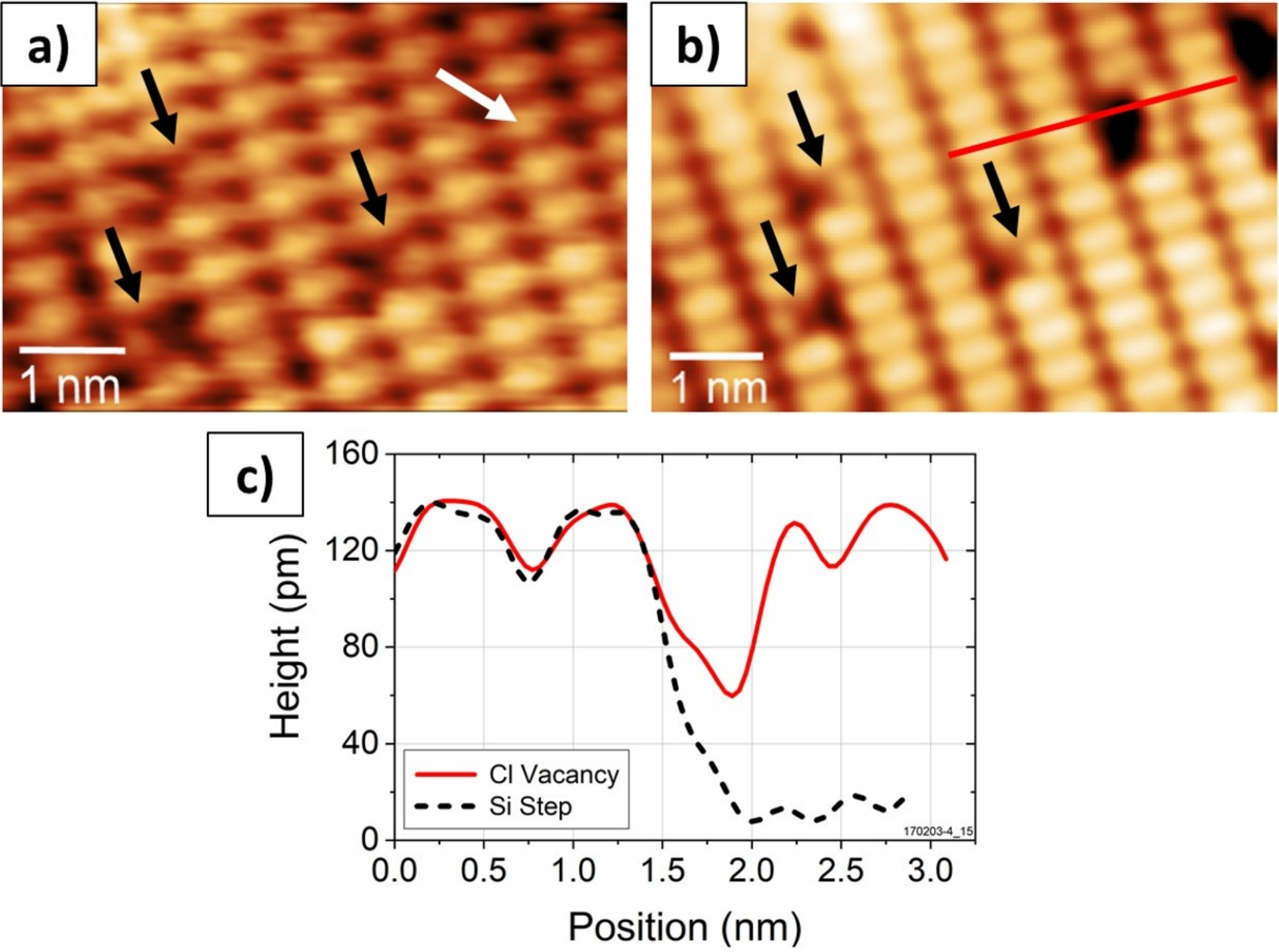}\hfill
\caption{Successive, filled-state STM images ($-$1.6 V, 0.5 nA) of a Cl-Si(100) surface at room temperature (a) before and (b) after a \mbox{500 $\mu$s} voltage pulse of \mbox{$+$4.25 V} was applied to the sample. $I_{t}$ was set to \mbox{0.5 nA} prior to the pulse with $V_{B} =$ \mbox{$-$1.6 V}. Three black arrows identify defects that serve as reference points between images. The white arrow in (a) indicates the location of the tip during the voltage pulse. (b) Immediately after the pulse, two single atom-sized defects are observed on neighboring Si dimers along the same dimer row. (c) Line profiles obtained along the line in (b) crossing the defect (solid line), and across a nearby single atomic-height step (dashed line) within the same data set (not shown). The apparent depth of the defect is shallower than a Si atomic step at about \mbox{0.8 {\AA}}, indicative of a Cl vacancy.}
\label{pos_pulse}
\end{figure}

Cl was selectively desorbed from the surface using voltage pulses between the tip and sample. Figure~\ref{pos_pulse} depicts successive filled-state STM images of the Cl-Si(100) surface of a lower doped substrate at room temperature (a) before and (b) after a \mbox{500 $\mu$s} voltage pulse of \mbox{$+$4.25 V} was applied to the sample. Prior to the pulse, the tunneling current setpoint, $I_{t}$, was set to \mbox{0.5 nA} with $V_{B} =$ \mbox{$-$1.6 V} and the STM feedback was disabled. The black arrows identify three defects that remained stationary during successive imaging and serve as points of reference. The voltage pulse was executed at the Cl-terminated dimer marked with the white arrow in Fig.~\ref{pos_pulse}(a). Immediately following the pulse, two single atom-sized vacancies appeared on neighboring Si dimers along the same dimer row, as shown in Fig.~\ref{pos_pulse}(b). Imaging also improved in sharpness from the elimination of a slight double-tip imaging artifact that was potentially due to the transfer of Cl to the tip, similar to H transfer\cite{Labidi2017, Huff2017}. Figure~\ref{pos_pulse}(c) displays a line profile of the defect (solid line) taken along the solid line in Fig.~\ref{pos_pulse}(b), which is compared to the line profile of a nearby single atomic-height step (dashed line) within the same data set (not shown). We found the apparent step height to be approximately \mbox{1.3 {\AA}}, close to the expected value of \mbox{1.36 {\AA}}. After adjusting for this height calibration error, we found the apparent depth of the vacancy structure created by the voltage pulse to be approximately \mbox{0.83 {\AA}}, less than that of a Si step. Additionally, the configuration of the created vacancy structures correspond well with those reported for halogen vacancies located on neighboring Si dimers\cite{Koji_PRL:2002, Trenhaile:2006}. We conclude that the $V_{B}$ pulse generated multiple Cl-atom vacancies and that Si was not removed from the site.

The vacancy structure created in Fig.~\ref{pos_pulse} is highly suggestive of a recombinative Cl$_{2}$ desorption process stimulated by the electrons injected from the STM tip. The arrangement of the Cl vacancies on neighboring Si dimers along the same dimer row is similar to that observed for the recombinative desorption of H$_{2}$ from H-Si(100)\cite{Durr:2002}. Cl$_{2}$ does not typically desorb from the surface by thermal, photon, or electron stimulation, and, as such, the desorption of Cl$_{2}$ from Cl-Si(100) has not been fully studied\cite{Kota:1998}. Instead, Cl is removed from the surface as SiCl$_{2}$ and SiCl$_{4}$ via etching processes\cite{GAO1993140, ALDAO2001189, JACKMAN1989296} and, to a lesser extent, as atomic Cl through a phonon-activated, electron-stimulated desorption process that is active near one monolayer halogen coverages\cite{TRENHAILE2005L135, Trenhaile:2006}. While we can conclude that Cl vacancies were created by the $V_{B}$ pulse, further investigation is required to determine the specific reaction mechanisms.

\begin{figure}[t!]
\centering
\includegraphics[width=\columnwidth]{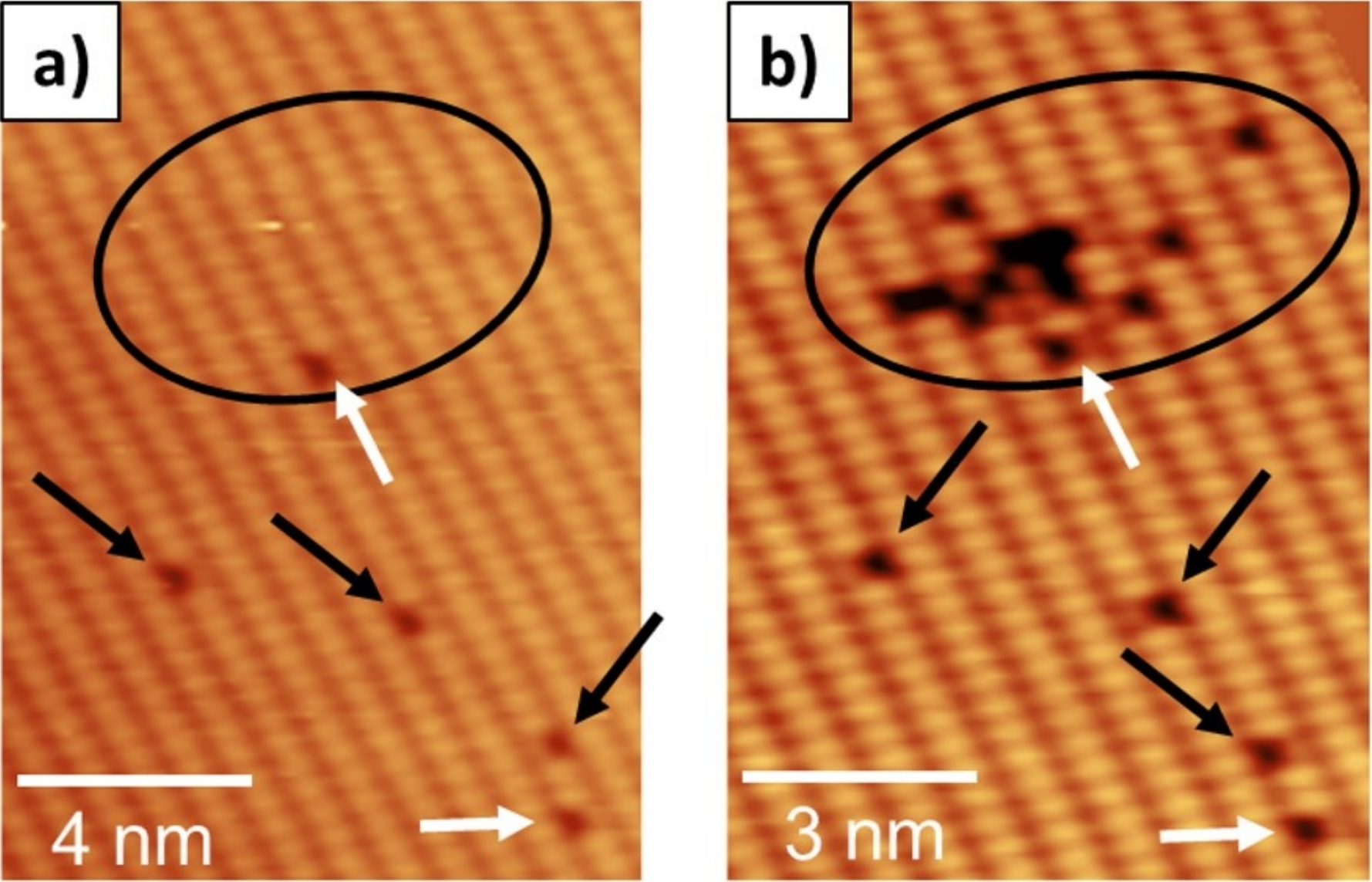}\hfill
\caption{Successive, filled-state STM images ($-$1.0 V, 0.8 nA) of a Cl-Si(100) surface at \mbox{600 K} (a) before and (b) after a \mbox{500 $\mu$s} voltage pulse of \mbox{$-$5.5 V} was applied to the sample. $I_{t}$ was set to \mbox{0.1 nA} with $V_{B} =$ \mbox{$-$1.0 V} prior to the pulse. Black arrows identify three single Cl vacancy defects that undergo intradimer hopping during imaging, while the white arrows identify similar defects that remained stationary. A total of 13 Cl vacancies were created by the pulse consisting of single, isolated vacancies as well as paired vacancies on Si dimers.}
\label{neg_pulse}
\end{figure}

In addition to positive $V_{B}$ pulses, in which desorption is facilitated via electron injection, we found that negative $V_{B}$ pulses, which correspond to hole injection, also induced desorption. Figure~\ref{neg_pulse} shows successive filled-state STM images (a) before and (b) after a \mbox{500 $\mu$s} voltage pulse of \mbox{$-$5.5 V} was applied to the Cl-Si(100) surface of a lower doped substrate at \mbox{600 K}. $I_{t}$ was set to \mbox{0.1 nA} with $V_{B} =$ \mbox{$-$1.0 V} prior to the pulse. Black arrows identify three defects initially present on the surface, which serve as reference points. In contrast to the reference features identified in Fig.~\ref{pos_pulse} (black arrows), we observed these features undergoing intradimer hopping from one side of the Si dimer to the other in subsequent images at \mbox{600 K}. This hopping is apparent when comparing defect positions in Fig.~\ref{neg_pulse}(a) and (b). The white arrows identify similar features that appear in the same position in both images, although they may still be hopping in between images. While thermal drift at \mbox{600 K} prevents us from knowing the exact pulse location, we observed a cluster of vacancies extending out roughly \mbox{3 nm} from the presumed pulse site located at the highest concentration of vacancies in Fig.~\ref{neg_pulse}(b). This negative $V_{B}$ pulse created a total of 13 Cl vacancies consisting of a mixture of single, isolated Cl vacancies as well as Cl di-vacancies, i.e., bare Si dimers. We observed similar results from comparable pulses, with many vacancies being created over a span of several nanometers.

The spatial extent of the stimulated reactions observed in Fig.~\ref{neg_pulse}(b) is in qualitative agreement with similar results of non-local atomic manipulation of adsorbates on Si(111) via hole injection\cite{Rusimova:2016, Lock_NCOMMS:2015, Sloan_PRL:2010}. While the reactions reported in Ref.~\citenum{Rusimova:2016} extend out tens of nanometers, we find hole injection induces reactions over a shorter length scale, presumably due to the much shorter pulse times reported here. We do not believe that the spatial extent of the vacancy complexes are due to outward vacancy diffusion since the neighboring single-atom vacancies remain stationary. If there was sufficient energy to drive outward diffusion, there would be sufficient energy to enable neighboring vacancies to pair in a more stable configuration on a single dimer, effectively gaining \mbox{0.4 eV} in energy due to the formation of a Si-Si $\pi$ bond\cite{Boland_PRL:1991}. Again, we are unable to definitively conclude from our results whether or not the negative voltage pulse induced the desorption of single Cl atoms, Cl$_{2}$ molecules, or a mixture of both, but the creation of an odd number of vacancies is suggestive of an active single-atom desorption mechanism.

Although STM-induced desorption of single H atoms has been demonstrated\cite{Moller:2017, Hersam:2000}, it was presumed to be too energetically costly to desorb Cl atoms from the surface due to the larger bonding energy of Si-Cl (\mbox{4.7 eV}) compared to Si-H (\mbox{3.1 eV})\cite{Dean1999}. Single Cl atoms can also excite into one of several metastable adsorption sites, such as the dimer bridge bond or a Si-Si back-bond, to produce inserted Cl atoms, Cl(i), and a dangling bond on the surface\cite{de_Wijs:1996, de_Wijs:1998, Boland1703}. Pairs of Cl(i) have a clearly defined STM signature, appearing as ``bright'' features extending across multiple dimer rows\cite{Agrawal:2007, Aldao:2009}. The absence of these bright features in our data suggests that the creation of dangling bonds is due to Cl desorption and not conversion to Cl(i). However, we cannot completely rule out the latter since the low diffusion barrier (\mbox{0.3 eV} to \mbox{0.4 eV}) from one inserted site to another\cite{de_Wijs:1998} is more easily overcome at these elevated temperatures, enabling Cl(i) to reach a more stable adsorption site at a step\cite{Butera:2009}.

\begin{figure}[t!]
\centering
\includegraphics[width=\columnwidth]{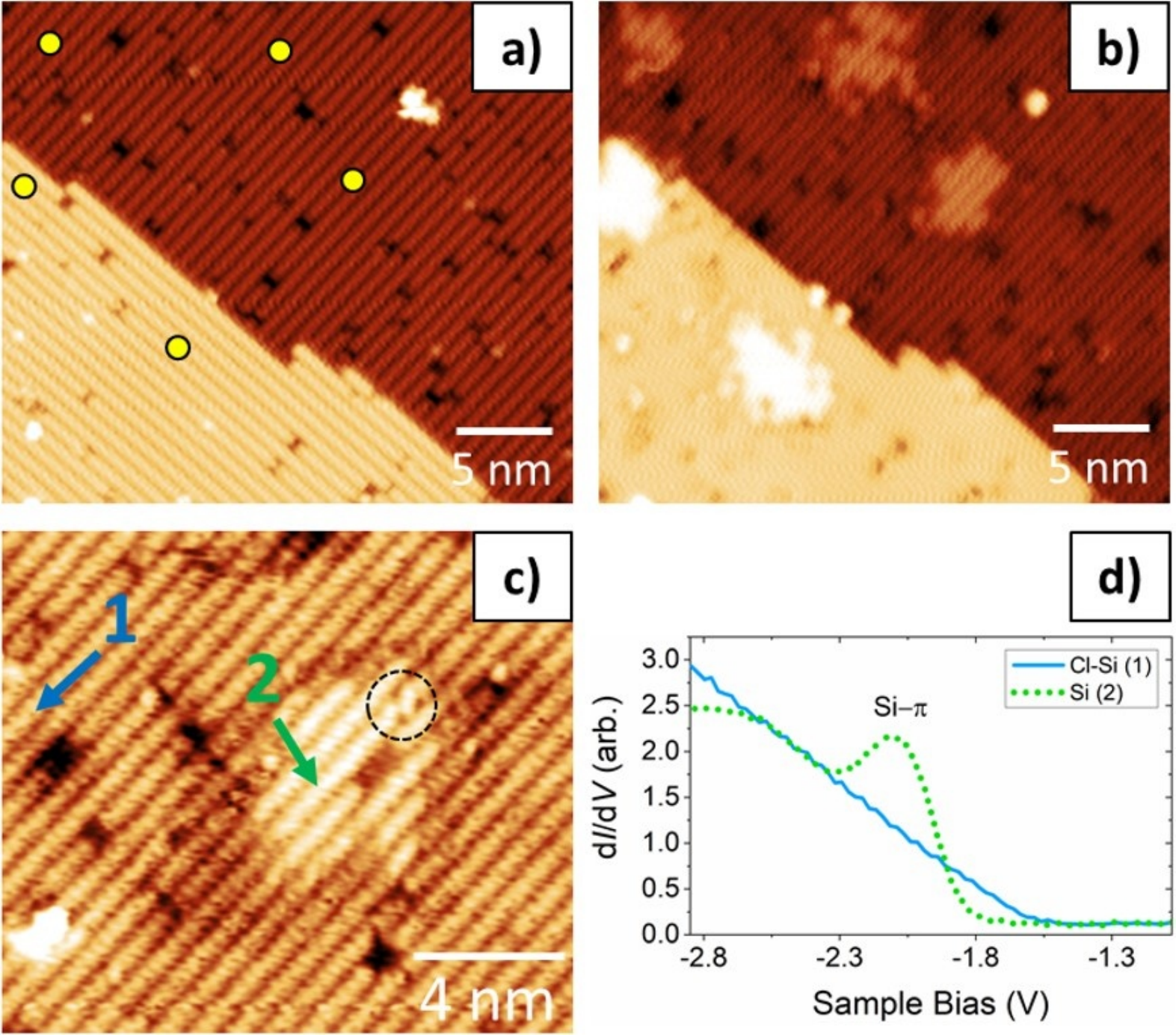}\hfill
\caption{Filled-state STM images ((a): $-$3.5 V, 0.02 nA, (b): \mbox{$-$3.5 V}, \mbox{0.03 nA}, (c): \mbox{$-$2.5 V}, \mbox{0.03 nA}) of Cl-Si(100) surfaces at \mbox{4 K}. (a) A fully Cl-terminated surface before a $V_{B}$ ramp is shown with a single atomic step running across the image. $V_{B}$ was ramped from \mbox{$-$3.5 V} to \mbox{$+$10.0 V} in roughly \mbox{0.5 s} with a maximum read current of several nanoamperes at five locations (dots) in (a). (b) After the $V_{B}$ ramps, five bright patches of Cl-free, bare dimers are visible. (c) Another patch of bare Si with buckled dimers (dashed circle) and several darker dimers visible within the bright area. (d) Averaged d$I$/d$V$ measurements taken on a Cl-terminated area (arrow 1 in (c)) and the bright, bare Si area (arrow 2 in (c)). The peak at \mbox{$-$2 V} in the spectrum from the bright area (dotted line) results from the Si $\pi$ bond of bare dimers. No $\pi$ bond peak is present for the Cl-terminated area (solid line).}
\label{cryo}
\end{figure}

Further evidence of the absence of Cl(i) on the surface was obtained at \mbox{4 K} where we expect Cl(i) diffusion to be frozen out. Additionally, Cl desorption was speculated to be more efficient at low temperature, as is the case for H-Si(100)\cite{Foley1998}. Figure~\ref{cryo} shows several filled-state images of Cl-Si(100) surfaces at a sample temperature of \mbox{4 K}. These were obtained on high-purity Si epilayer substrates using the cryogenic STM system\cite{our_4K_system}. Figure~\ref{cryo}(a) and (b) are successive images taken before and after $V_{B}$ was ramped from \mbox{$-$3.5 V} to \mbox{$+$10.0 V} and back with $I_{t} =$ \mbox{0.03 nA} at the five locations marked by dots in (a). The ramp time was roughly \mbox{1 s} and the max current read several nanoamperes. These $V_{B}$ ramps resulted in five bright areas on the surface corresponding to Cl-free, bare Si dimers. The voltage threshold for desorption was found to be higher at low temperature since no desorption was observed for $V_{B}$ ramps less than \mbox{$+$10.0 V}. STM spectroscopy was used to verify the surface termination of a bright area shown in Fig.~\ref{cryo}(c), which was created using a $V_{B}$ ramp similar to the one used in Fig.~\ref{cryo}(b). Figure~\ref{cryo}(d) shows averaged d$I/$d$V$ measurements taken from a presumed Cl-terminated area (arrow 1 in (c)) as well as the bright area (arrow 2 in (c)). As seen in Fig.~\ref{cryo}(d), a peak at approximately \mbox{$-$2 V} is present in the d$I/$d$V$ spectrum from the bright area (dotted line), indicative of the Si $\pi$ bond formed on the bare Si dimer\cite{PhysRevB.73.035330}. The position of this peak at \mbox{$-$2 V} is due to the low doping of the substrate. The spectrum from the Cl-terminated area (solid line) is linear and shows no $\pi$-bond peak, verifying that the bright area was bare Si while the surrounding darker areas remained Cl terminated. Further, buckled dimers visible within the dashed circle in the bright area are also indicative of bare Si. At \mbox{4 K}, we did not observe any evidence of nearby Cl(i), and thus we conclude that Cl was completely removed from the surface and not converted to Cl(i).

The bright appearance of the bare Si dimers seen in Fig.~\ref{cryo}(b) and (c) is in contrast to the dark appearance of the Cl vacancy features in Fig.~\ref{pos_pulse} and Fig.~\ref{neg_pulse}. Unlike the H-Si(100) surface, where H vacancies (bare Si) appear bright compared to surrounding H-terminated areas, Cl vacancies can appear either dark or bright depending on scanning conditions and whether the defects are single vacancies (typically dark in filled-state imaging) or di-vacancies, i.e., bare dimers (typically bright in filled-state imaging)\citep{Lee_BD:2004}. Sample temperature may also play a role in imaging these defects. In Fig.~\ref{cryo}(b) and (c), multiple bare dimers are aligned together on several dimer rows resulting in the bright appearance of the desorbed patches, while the Cl vacancy cluster in Fig.~\ref{neg_pulse}(b) consists mostly of single vacancies.

\begin{figure}[t!]
\centering
\includegraphics[width=\columnwidth]{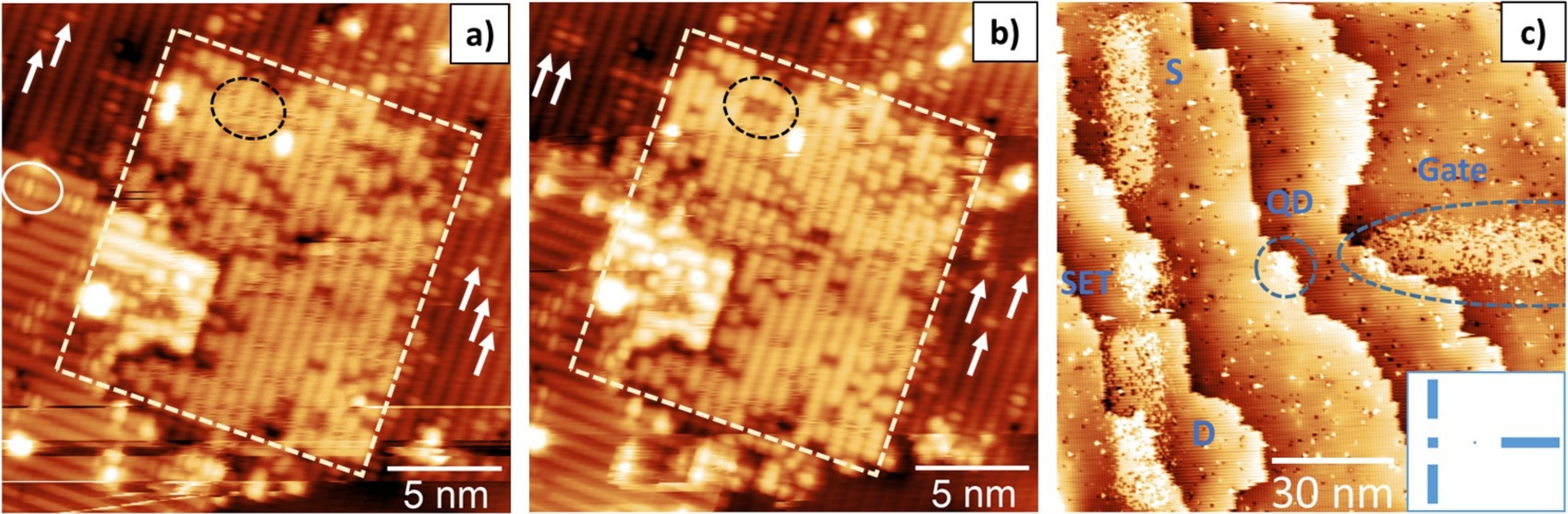}\hfill
\caption{Filled-state STM images ((a) and (b): \mbox{$-$1.0 V}, \mbox{0.8 nA}, (c): \mbox{$-$1.6 V}, \mbox{0.7 nA}) of Cl-Si(100) surfaces at \mbox{400 K} showing lithographic patterns created with positive $V_{B}$ values in a field emission patterning mode. (a) and (b) Successive images of a square patterned in a serpentine manner along every other dimer row with $V_{B} =$ \mbox{$+$9.0 V}, $I_{t} =$ \mbox{1 nA}, and $d_{e} =$ \mbox{10 mC/cm}. Cl-free Si dimers appear bright. Dark, Cl-terminated dimers are visible within the square. A dashed box outlining the pattern shows that the dimensions did not change over multiple scans. Spurious bare dimers are visible outside the patterned area as bright triplet features (solid oval). White arrows in (a) and (b) highlight the diffusion of Cl di-vacancies along dimer rows at \mbox{400 K}. Intradimer hopping of remaining Cl within the patterned area appears as blurred dimers (dashed oval) in (a). Dark, Cl-terminated dimers (dashed oval) resulting from the pairing of blurred dimers are seen in (b). (c) A larger area image showing a patterned mock-up of a quantum device created using parameters similar to those in (a). It includes a quantum dot island (QD), an electrostatic gate, and a single electron transistor (SET) with source (S) and drain (D) gates. The inset shows the intended pattern.}
\label{FE-squares}
\end{figure}

In order for Cl to be a viable resist for device fabrication via STM lithography, it is necessary to pattern relatively large areas and remove a sufficient amount of Cl to enable adsorption of the precursor molecule. Lithographic patterning of H-Si(100) is routinely done at room temperature; however, we found patterning Cl-Si(100) to be more facile at elevated temperatures. This is likely due to the larger bonding energy of Si-Cl compared to Si-H. Figure~\ref{FE-squares} shows filled-state images demonstrating STM lithography on Cl-Si(100) in a field emission patterning mode utilizing a positive $V_{B}$. This lithography was done while the sample was at a temperature of \mbox{400 K} on a higher doped substrate. Figure~\ref{FE-squares}(a) and (b) are successive filled-state STM images after lithographically patterning a \mbox{10 nm} by \mbox{10 nm} square area. The square was produced with $V_{B} =$ \mbox{$+$9.0 V}, $I_{t} =$ \mbox{1 nA}, and $d_{e} =$ \mbox{10 mC/cm} (write speed of \mbox{1 nm/s}) while the STM tip moved along every other dimer row of the patterned area in a serpentine manner\cite{Zyvex_HDL:2014}. The electron dose, $d_{e}$, used here represents the total number of electrons delivered per unit length along the line (units of mC/cm) and sets the write speed for a given $I_{t}$ setpoint value.

Here, as in Fig.~\ref{cryo}, Cl-free dimers appear bright. Remnant Cl atoms found within the patterned area from incomplete depassivation tend to pair, forming fully Cl-terminated dimers with a darkened appearance. The pattern is continuous across the single atomic-height step located on the left-hand side of the image. The dashed box roughly outlines the patterned area serving as a marker for pattern fidelity. Figure~\ref{FE-squares}(b) shows that over several successive scans, the general dimensions of the pattern did not change. Diffusion of adsorbed Cl atoms into the patterned area from the surrounding edges is not significant at these slightly elevated temperatures. Numerous spurious bare Si dimers are visible outside the patterned area extending several nanometers away from the pattern edges. While positive $V_{B}$ field emission patterning is relatively quick and can cover a large area, it suffers from stochastic desorption events relatively far from the tip location. The desorption of Cl outside of the intended area is akin to spurious H desorption observed during STM patterning of H-Si(100)\cite{Zyvex_Spurious:2014}. Isolated Cl di-vacancies, such as the one highlighted by the solid oval in Fig.~\ref{FE-squares}(a), appear as a bright triplet feature at modest scan biases\cite{Koji_PRL:2002, Lee_BD:2004}. While the main pattern remained intact at elevated temperatures, individual Cl di-vacancies were observed diffusing along the dimer rows outside of the pattern, as indicated by the white arrows in Fig.~\ref{FE-squares}(a) and (b).

At \mbox{400 K}, we also observed intradimer hopping of Cl atoms within patterned areas. Intradimer hopping is imaged by the STM as a blurred feature with lobes at both ends of the Si dimer. It is the result of a single Cl atom hopping from one side of a Si dimer to the other at a rate much faster than the image acquisition time. The barrier for intradimer hopping was previously found to be approximately \mbox{0.6 eV}\cite{de_Wijs:1996} and this process was shown to be facilitated by an STM tip under normal imaging conditions\cite{NAKAMURA200368}. In contrast, the activation barrier for interdimer diffusion is \mbox{1.1 eV}\cite{de_Wijs:1996}. We did observe the pairing of several neighboring Cl atoms, visible as the blurred dimers in the dashed oval in Fig.~\ref{FE-squares}(a), to create fully Cl-terminated dimers, appearing dark within the dashed oval in Fig.~\ref{FE-squares}(b). The decay of a Cl-terminated dimer into two neighboring Cl atoms undergoing intradimer hopping was also observed.

Figure~\ref{FE-squares}(c) shows field emission patterning in a larger area. A mock-up of a quantum device including a quantum dot island (QD), an electrostatic gate, and a single electron transistor (SET) charge sensor was patterned. The inset shows the intended pattern. This pattern was created using similar parameters as those used in Fig.~\ref{FE-squares}(a).

\begin{figure}[t!]
\centering
\includegraphics[width=0.33\textwidth]{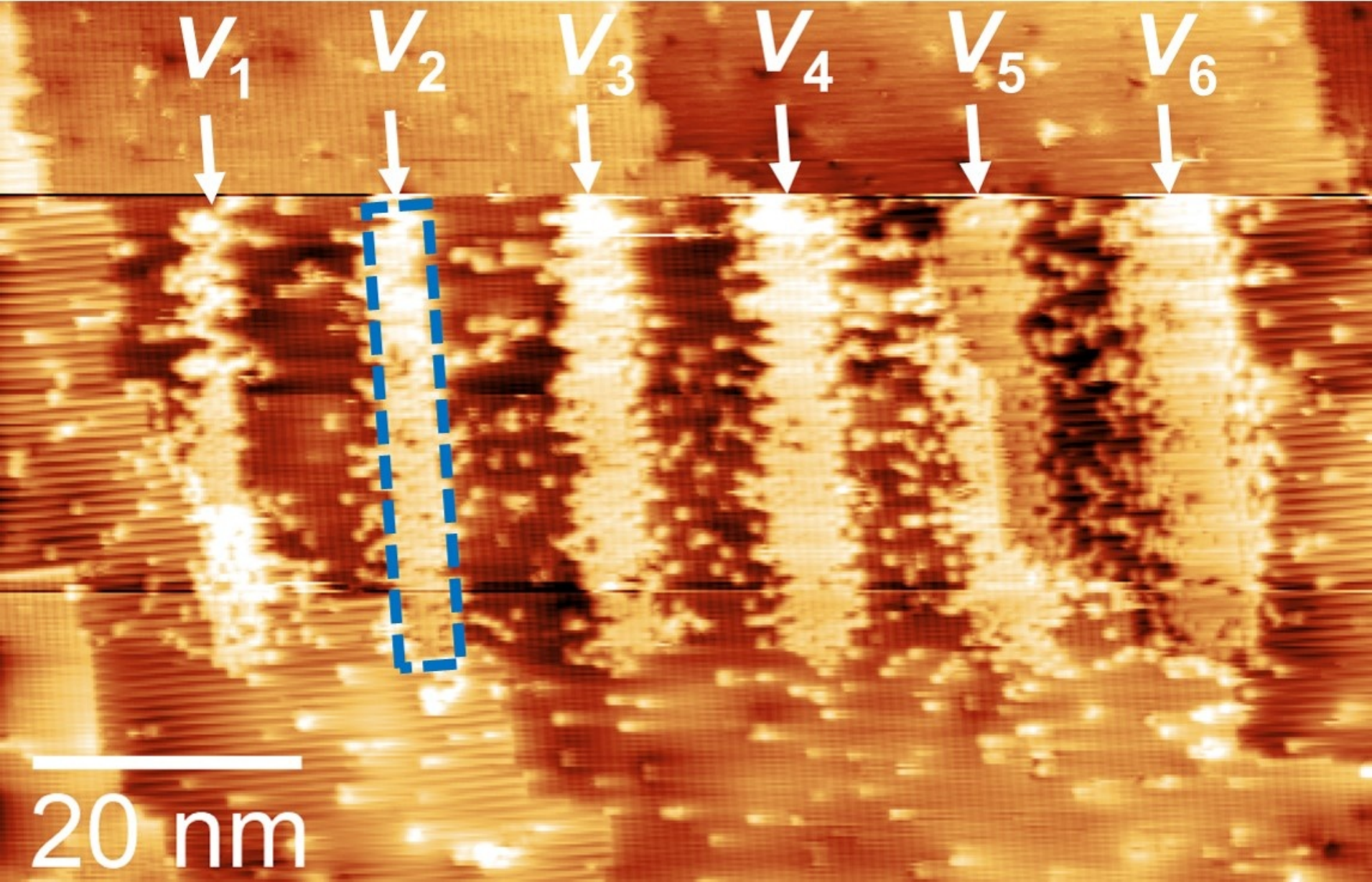}\hfill
\caption{Filled-state STM image ($-$1.5 V, 0.7 nA) of lines lithographically patterned on a Cl-Si(100) surface at \mbox{400 K} using field emission patterning. The lines, indicated by the arrows, are clearly continuous across steps in the Si surface. Each line was written using positive $V_{B}$ values between $V_{1} =$ \mbox{$+$7.5 V} and $V_{6} =$ \mbox{$+$10.0 V}, increasing from left to right in steps of \mbox{0.5 V}. For all lines, $I_{t} =$ \mbox{5 nA} and $d_{e} =$ \mbox{15 mC/cm}. The widths of the depassivated lines increase with increasing $V_{B}$. The measured width of the line patterned with bias $V_{2}$ is represented by the dashed box.}
\label{parm-search}
\end{figure}

In an effort to better understand and optimize field emission lithography of Cl, we patterned sets of lines while independently varying $V_{B}$, $I_{t}$, and $d_{e}$ and compared the resulting pattern line widths. Figure~\ref{parm-search} is a filled-state image depicting a set of six lines lithographically patterned with varying $V_{B}$ values at \mbox{400 K} on a higher doped substrate. Several of the lines can be seen going over the single atomic steps in the surface. Each line was written with $I_{t} =$ \mbox{5 nA} and $d_{e} =$ \mbox{15 mC/cm} while varying $V_{B}$ from $V_{1} =$ \mbox{$+$7.5 V} to $V_{6} =$ \mbox{$+$10.0 V} in steps of \mbox{0.5 V}. Qualitatively, the widths of the lines in Fig.~\ref{parm-search} appear larger at higher $V_{B}$ values. Quantitatively, the line widths were measured by averaging the profiles across a given line over the length of the line and then extracting the full width at half max of that averaged height profile. The measured width of the line patterned with bias $V_{2}$ in Fig.~\ref{parm-search} is represented by the dashed box around the pattern. The line widths increase from \mbox{4.6(5) nm} for $V_{B} =$ \mbox{$+$7.5 V} to \mbox{7.9(5) nm} for $V_{B} =$ \mbox{$+$10.0 V}.

\begin{figure}[th!]
\centering
\includegraphics[width=\columnwidth]{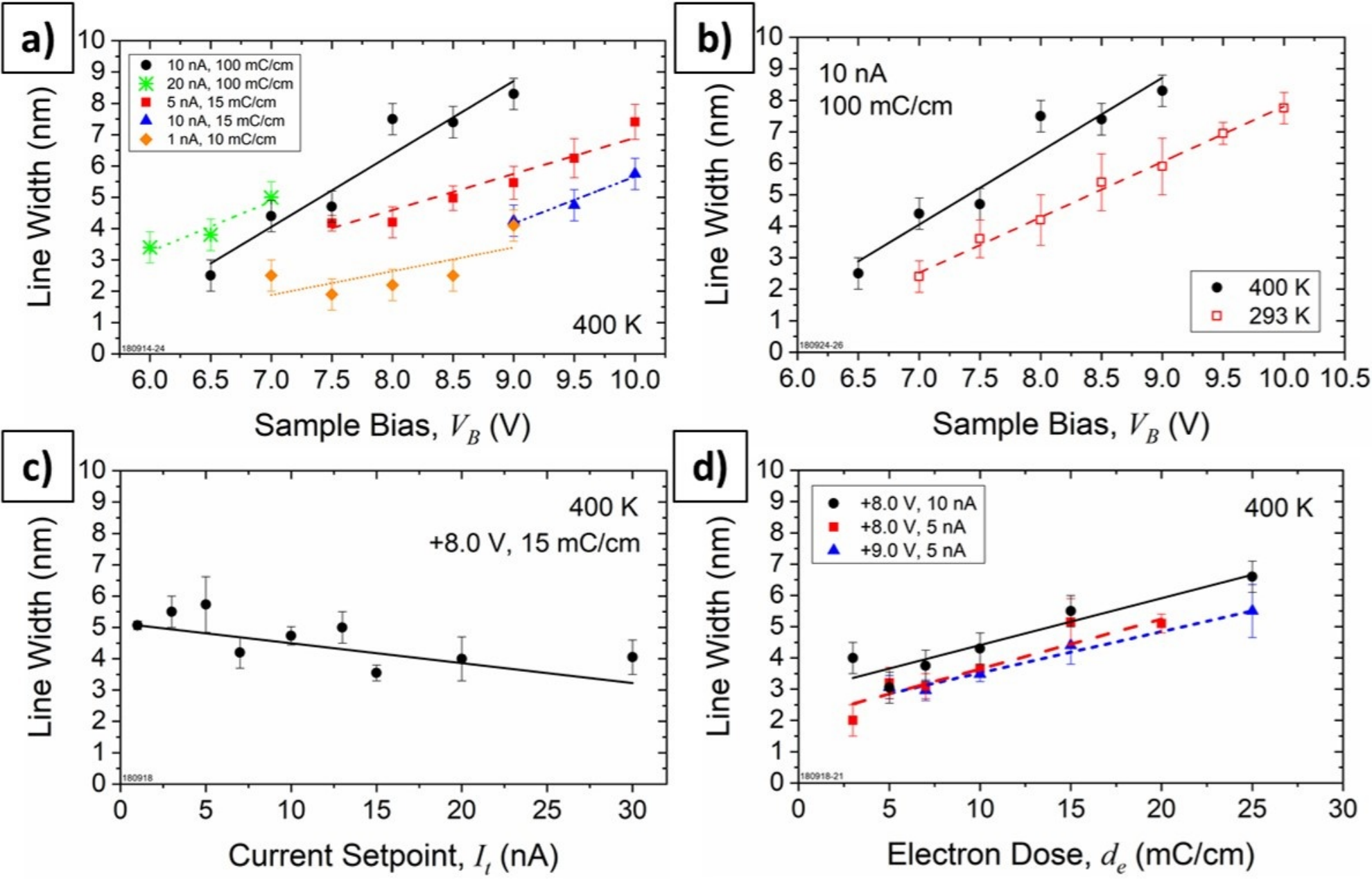}
\caption{Measured widths of lithographic lines patterned in Cl-Si(100) as a function of (a) positive $V_{B}$, (b) sample temperature, (c) $I_{t}$, and (d) $d_{e}$. Linear fits (lines) to each data set (symbols) are also shown. (a) Line widths increase linearly with increasing $V_{B}$ for lines patterned at \mbox{400 K} using multiple parameters. The average slope of the data in (a) is \mbox{1.7(2) nm/V}. (b) Widths of \mbox{400 K} lines (circles) are compared to those patterned at room temperature (open squares) with $I_{t} =$ \mbox{10 nA} and $d_{e} =$ \mbox{100 mC/cm} for both sets. Widths of \mbox{400 K} lines are shifted to lower $V_{B}$ values relative to those of room temperature lines. (c) Line widths weakly decrease with increasing $I_{t}$ for lines patterned at \mbox{400 K} with $V_{B} =$ \mbox{$+$8.0 V} and $d_{e} =$ \mbox{15 mC/cm}. (d) Line widths increase linearly with increasing $d_{e}$ for lines patterned at \mbox{400 K} using multiple parameters. The average slope of the data in (d) is \mbox{0.15(2) nm/(mC/cm)}.}
\label{sweeps}
\end{figure}

Figure~\ref{sweeps} displays the measured widths of several sets of lithographic lines patterned at \mbox{400 K} as a function of (a) positive $V_{B}$, (b) sample temperature, (c) $I_{t}$, and (d) $d_{e}$. Linear fits to each data set (lines) are also shown. The measured line widths in Fig.~\ref{sweeps}(a) are shown to increase linearly with $V_{B}$ across multiple lithographic parameters. $I_{t}$ ranged from \mbox{1 nA} to \mbox{20 nA} and $d_{e}$ ranged from \mbox{10 mC/cm} to \mbox{100 mC/cm}. The slopes of each linear fit are consistent across the data, although there are clear offsets between the data sets. Table~\ref{tbl:bias} displays the slopes and lithographic parameters for each data set in Fig.~\ref{sweeps}(a). The average slope of all the data sets in (a) is \mbox{1.7(2) nm/V}. In comparison, a value of \mbox{2.2 nm/V} was previously reported for H-terminated Si(100) depassivation\cite{Lyding1996}.

The positive correlation observed between line width and $V_{B}$ is consistent with the expectation that higher bias voltages cause the tip to retract slightly. Thus, electrons are field emitted from the tip in a wider angular distribution and with higher energies, leading to more desorption over a larger area. The offset of the different data sets appears to be partially correlated to the other parameters, with lower $d_{e}$ values resulting in smaller line widths (e.g., diamonds) for a given $V_{B}$ and vice versa. We also observed line width offsets between distinct data sets patterned using the same values of $I_{t}$ and $d_{e}$, however the slopes remained consistent. These offsets are likely due to the sharpness and condition of the tip as well as the patterning alignment with the dimer rows. Extrapolating the fit slopes back to zero line width gives an approximate measure of the threshold bias value where Cl atoms receive enough energy to desorb from the surface. This fit-derived threshold bias ranges from \mbox{$+$4.0 V} to \mbox{$+$7.5 V} with an average of \mbox{$+$5.4(4) V}. These values are consistent with the typical positive $V_{B}$ values of \mbox{$+$6.0 V} to \mbox{$+$6.5 V} we observed for the onset of depassivation on Cl-Si(100) at elevated temperatures.

Figure~\ref{sweeps}(b) compares the widths of \mbox{400 K} lines (circles) to those of lines patterned at room temperature (open squares). Both sets of lines were patterned with $I_{t} =$ \mbox{10 nA} and $d_{e} =$ \mbox{100 mC/cm}. The \mbox{400 K} line widths are shifted to lower $V_{B}$ values relative to the room temperature line widths. While a shift may be the result of factors unrelated to temperature, such as changes in the tip, we saw a consistent shift to lower $V_{B}$ values at \mbox{400 K} compared to room temperature. Moreover, at \mbox{4 K}, a minimum $V_{B}$ of \mbox{$+$10.0 V} was needed to initiate desorption, indicating that the thermal energy of Cl on the surface plays a roll in desorption.

Figure~\ref{sweeps}(c) shows that line width weakly decreases with increasing $I_{t}$. These data are the average of several sets of lines patterned at \mbox{400 K} with $V_{B} =$ \mbox{$+$8.0 V} and $d_{e} =$ \mbox{15 mC/cm}. The constant or weakly inverse correlation between line width and $I_{t}$ can be partly attributed to the tip moving closer to the surface as $I_{t}$ increases. This results in a narrower spread of emitted electrons. Additionally, while a higher current would presumably lead to more desorption events, a constant $d_{e}$ results in the write speed increasing at larger $I_{t}$ values, causing the tip to spend less time at each surface site.

Finally, Fig.~\ref{sweeps}(d) shows the line width dependence on $d_{e}$ for lines patterned at \mbox{400 K} using $V_{B}$ values of \mbox{$+$8.0 V} and \mbox{$+$9.0 V} with $I_{t}$ values of either \mbox{5 nA} or \mbox{10 nA}. Two of the data sets (squares and triangles) are averages of several sets of measured lines. The measured line widths increase linearly with increasing $d_{e}$. Both the line widths and the slopes of the fits are consistent across the data sets. Table~\ref{tbl:dose} displays the slopes and lithographic parameters for each data set in Fig.~\ref{sweeps}(d). The average slope of the data in (d) is \mbox{0.15(2) nm/(mC/cm)}. Extrapolating previously reported electron dose data for depassivation of H-terminated Si(100) to our measurement range gives a similar line width dependence of approximately \mbox{0.22 nm/(mC/nm)}\cite{Lyding1996}. The positive correlation between line width and $d_{e}$ is consistent with the fact that at larger $d_{e}$ values, more electrons are injected over the same area leading to more chances for desorption events. Additionally, we observed that higher $d_{e}$ values resulted in more complete depassivation along the lithographic line, i.e., fewer Cl-terminated dimers remaining. The average depassivation fraction (bare Si dimer area over the total area) along lines presented in Fig.~\ref{sweeps}(d) increases from 0.72(3) for $d_{e}$ values of \mbox{3 mC/cm} and \mbox{5 mC/cm} up to 0.84(1) for $d_{e}$ values of \mbox{20 mC/cm} and \mbox{25 mC/cm}. For $d_{e}$ $\leq$ \mbox{3 mC/cm}, we often observed little to no depassivation. For all the lines in Fig.~\ref{sweeps} patterned using positive $V_{B}$, the average depassivation fraction is 0.83(1), although depassivation increases away from the edges of lines and can be near unity around the center. While more complete depassivation may seem desirable, it is not necessary for lithographic features to be perfectly continuous to fabricate metallic regions via dopant incorporation. This is due to the large spatial extent of the dopant electronic states (\mbox{2.5 nm} Bohr radius for P), which can bridge the gap between narrow undoped regions\cite{Weber64}.

\begin{figure}[t!]
\centering
\includegraphics[width=\columnwidth]{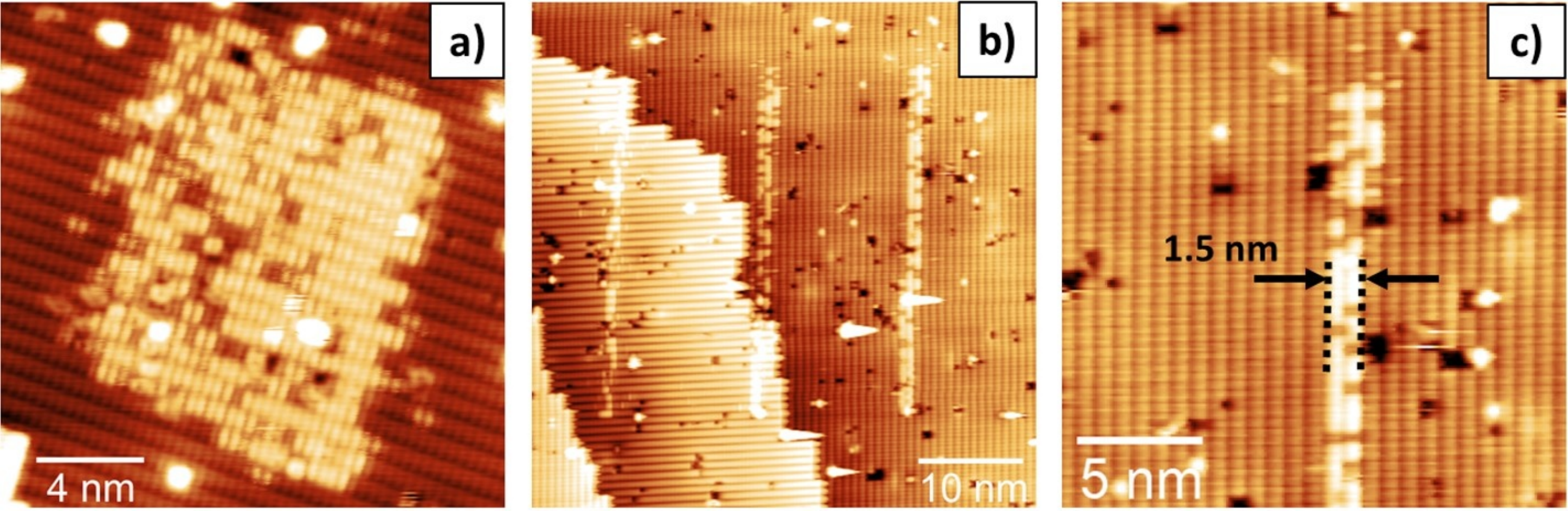}\hfill
\caption{Filled-state STM images ((a): \mbox{$-$1.7 V}, \mbox{0.8 nA}, (b) and (c): \mbox{$-$1.6 V}, \mbox{0.7 nA}) of Cl-Si(100) surfaces at \mbox{400 K} showing lithographic patterns of (a) a square and (b) and (c) lines produced using atomic precision patterning with negative $V_{B}$ values. Bright areas within the square and along the lines are Cl-free Si dimers. Dark, Cl-terminated dimers are also visible within the patterns. The square in (a) was patterned in a serpentine manner across dimer rows with $V_{B} =$ \mbox{$-$6.5 V}, $I_{t} =$ \mbox{10 nA}, and $d_{e} =$ \mbox{10 mC/cm}. The pattern has well-defined edges with few spurious bare dimers surrounding it. The lines in (b) were patterned with $V_{B} =$ \mbox{$-$6.5 V} and $d_{e} =$ \mbox{10 mC/cm}. The leftmost line was patterned with $I_{t} =$ \mbox{10 nA} and the right two with $I_{t} =$ \mbox{100 nA}. The center line is continuous across the atomic step running though the image. The widths of the patterned lines in (b) vary between one and three dimer rows. The average fraction of depassivated dimers for these lines is 0.66(4). (c) The rightmost line in (b) is \mbox {1.5 nm} wide with darker, Cl-terminated dimers visible along the line.}
\label{AP-litho}
\end{figure}

At the lowest positive $V_{B}$ values used for desorption, no atomically precise patterning was observed. In contrast, an atomic precision patterning mode on H-Si(100) is active at $V_{B}$ values below the Cl desorption threshold bias we observed (e.g., \mbox{$+$4.0 V}). Interest in more precise patterning suitable for atomic-scale device fabrication led us to explore lithography with negative $V_{B}$ values. Negative biases have been used to induce the desorption of H via hole resonances\cite{Stokbro_PRL:1998}, and we found that hole injection induced desorption of Cl in a more confined, precise, and controllable manner than electron injection.

Figure~\ref{AP-litho} shows filled-state images demonstrating lithography on Cl-Si(100) using negative $V_{B}$ values in an atomic precision patterning mode. Lithography was done at \mbox{400 K} on a higher doped substrate. A \mbox{15 nm} by \mbox{15 nm} square is shown in Fig.~\ref{AP-litho}(a). It was patterned with $V_{B} =$ \mbox{$-$6.5 V}, $I_{t} =$ \mbox{10 nA}, and $d_{e} =$ \mbox{10 mC/cm} while the tip traveled across dimer rows in a serpentine manner instead of along them as before. Due to incomplete depassivation, a few dark, Cl-terminated dimers within the patterned area can be seen forming some extended structures. The reduced desorption yield is due to the tip having moved across dimer rows during the specific patterning procedure used here, which we found to be less effective than moving along dimer rows. Interestingly, there are significantly fewer spurious bare dimers in the surrounding areas and the edges of the pattern are more abrupt and well defined than with positive $V_{B}$ values.

As a demonstration of atomic precision patterning, Fig.~\ref{AP-litho}(b) shows three lines that vary in width from one to two dimer rows over most of the length of the lines, i.e, roughly \mbox{1.5 nm} or less. Each is \mbox{25 nm} in length and was patterned with $V_{B} =$ \mbox{$-$6.5 V} and $d_{e} =$ \mbox{10 mC/cm}. The line on the left was patterned with $I_{t} =$ \mbox{10 nA} while the center and right lines were patterned with $I_{t} =$ \mbox{100 nA}. The center line was easily patterned across the atomic step running vertically through the middle of the image. The rightmost line and most of the center line were patterned with the write direction roughly aligned parallel with the dimer rows. A slight angular misalignment caused the pattern to drift one dimer row over in a couple spots causing the lines to be three dimer rows wide in these areas. In comparison, we found that the thinnest lines written with positive $V_{B}$ values had widths spanning three or four dimer rows with more inconsistent depassivation.

We found little to no spurious bare dimers appearing away from the lines patterned along the dimer rows. In contrast, the leftmost line was patterned with the write direction perpendicular to the dimer rows due to the presence of the step. In this case, spurious bare dimers appeared several nanometers on either side of the lines, as can be seen upon close examination of the leftmost line in Fig.~\ref{AP-litho}(b). The narrowness of these lines is also clearly displayed in Fig.~\ref{AP-litho}(c), which depicts the rightmost line from Fig.~\ref{AP-litho}(b). The lines here are not fully continuous due to Cl remaining along the path of the lithography, which appear as several darker, Cl-terminated dimers. The average depassivation fraction of Si dimers along these three lines is 0.66(4).

Lithographic parameters for this atomic precision, negative $V_{B}$ mode were varied in a manner similar to the studies represented in Fig.~\ref{sweeps}, however, we found inconsistent or no dependence of the line widths on $V_{B}$, $I_{t}$, or $d_{e}$. Additionally, no dependence on those lithographic parameters was found for the depassivation fraction along the lines. The average line width across all of the patterned lines studied here was measured to be \mbox{1.74(7) nm}, which corresponds to 2.26(9) dimer row widths. The average depassivation fraction for all of the patterned lines produced using the range of lithographic parameters explored here was 0.68(2), lower than that of positive $V_{B}$ lines (0.83(1)). Ultimately, we found that the STM tip played a major role in the patterning precision when using negative $V_{B}$ values. Further studies are needed to better explore the lithography parameter phase space for atomic precision patterning, but it is clear that negative $V_{B}$ patterning is robust within a wide range of lithography conditions making it a highly valuable tool for fabrication of atomic scale devices on Cl-Si(100).

\section{Conclusion}
We have demonstrated STM-induced desorption and lithographic patterning of Cl-terminated Si(100)-(2$\times$1) from \mbox{4 K} to \mbox{600 K}. Elevated temperatures were found to aid in the Cl desorption process. Lithographically defined patterns were relatively stable up to \mbox{600 K}, potentially enabling selective area deposition at elevated temperatures. We utilized positive $V_{B}$ values for field emission patterning useful for patterning large areas and demonstrated atomic precision patterning with negative $V_{B}$ values where Cl is desorbed from one to two dimer rows. Varying positive $V_{B}$ lithographic patterning parameters revealed that line widths are positively correlated with $V_{B}$ and $d_{e}$, while the feature sizes of negative $V_{B}$ patterning were largely independent of these parameters. The results presented here clearly demonstrate the viability of Cl as a patternable resist on Si(100), furthering the development of halogen-based chemistries for atomic-precision, advanced-manufacturing applications.

We expect similar STM patterning behavior on bromine- and iodine-terminated Si(100)-(2$\times$1), and investigations are underway to explore these heavier halogens as patternable resists. Both Br and I are less tightly bound to Si than Cl and, as such, have the potential to result in a more efficient patterning process\cite{XU200577}. I is mobile at room temperature on Si(100)\cite{XU200577}, but at half-monolayer coverage, it is locked into a self-assembled c(4$\times$2) pattern\cite{Xu:2005} with every other Si dimer I-terminated. These self-assembled patterns have the potential to serve as templates for targeted adsorption into a periodic array on H-Si(100)\cite{Ferng:2009} or Cl-Si(100). We continue to explore the phase space of lithographic parameters to enable patterning of complex designs such as atomic-scale devices utilizing halogen-based acceptor chemistries such as BCl$_{3}$ or AlCl$_{3}$.

\section{Experimental Methods}
The experiments were performed in two separate UHV STM systems. Room temperature (approximately \mbox{293 K}) and elevated temperature experiments up to \mbox{600 K} were performed in a UHV chamber with a base pressure of $P$ \mbox{$< 6.7 \times 10^{-9}$ Pa} (\mbox{$5 \times 10^{-11}$ Torr}) using a ScientaOmicron VT-STM with a ZyVector STM Lithography control system. This STM allows for a maximum bias voltage, $V_{B}$, of $\pm$ \mbox{10.0 V} to be applied to the sample stage. The Si(100) wafers used were p-type, B-doped with either $\rho =$ \mbox{0.001 $\Omega\cdot$cm} to $\rho =$ \mbox{0.002 $\Omega\cdot$cm} or $\rho =$ \mbox{1 $\Omega\cdot$cm} to $\rho =$ \mbox{10 $\Omega\cdot$cm} and were oriented within 0.5$^{\circ}$ of (100). Si samples were cut to \mbox{4 mm} by \mbox{12 mm} in size, mounted on a ScientaOmicron XA sample plate, and loaded into the UHV chamber. Clean Si(100)-(2$\times$1) surfaces were prepared by flash annealing them in UHV to \mbox{1200 $^{\circ}$C}, following the procedure found in Ref.~\citenum{Trenhaile_O:2006}. Shortly after clean surface preparation, the samples were exposed to a flux of Cl$_{2}$ to achieve a Cl-terminated surface. Cl$_{2}$ was generated from a solid-state, electrochemical cell consisting of AgCl doped with 5 wt \% CdCl$_{2}$\cite{Spencer:1983}.
\begin{table}[b!]
  \caption{Lithographic patterning parameters including current setpoint, $I_{t}$, and electron dose, $d_{e}$, as well as linear fit slopes, $m$, for sets of lines patterned as a function of sample bias, $V_{B}$, in Fig.~\ref{sweeps}(a).}
  \label{tbl:bias}
  \begin{tabular}{llll}
    \hline
    data set                              & $I_{t}$ (nA) & $d_{e}$ (mC/cm) & $m$ (nm/V) \\
    \hline
    circles                               & 10           & 100             & 2.3(3)     \\
    stars                                 & 20           & 100             & 1.6(5)     \\
    squares                               & 5            & 15              & 1.2(2)     \\
    triangles                             & 10           & 15              & 1.5(3)     \\
    diamonds                              & 1            & 10              & 0.8(4)     \\
    H-Si(100) (Ref.~\citenum{Lyding1996}) & 0.1          & 0.54            & 2.2        \\
    \hline
  \end{tabular}
\end{table}
Cl exposures were done with the sample at roughly \mbox{600 K} to minimize water contamination and limit Cl atom insertion into the surface\cite{Yu:2008, Trenhaile_O:2006, Butera:2009, Aldao:2009}. This temperature was sufficient to ensure that any Cl(i) generated as a result of the chemisorption process can diffuse and attach at a dangling bond\cite{Agrawal:2007, Aldao:2009, Butera:2009}, but low enough to prevent activation of etching and roughening processes\cite{Koji_ROUGH:2002,ALDAO2001189}. During Cl$_{2}$ exposure, the pressure in the chamber remained below \mbox{$9.3 \times 10^{-9}$ Pa} (\mbox{$7 \times 10^{-11}$ Torr}).

The elevated sample temperatures on the STM stage were controlled by a LakeShore 335 temperature controller and monitored by means of a Pt-100 resistor mounted on the stage. Samples were left to equilibrate for at least one hour after each temperature change prior to imaging and lithography to minimize thermally induced drift. Lithographic patterning of the Cl-Si(100) surfaces was done by defining $V_{B}$, $I_{t}$, and $d_{e}$ within ZyVector scrips that generate and execute box and line patterns aligned to the Si(100) dimer rows. Voltage pulsing was executed with STM feedback turned off after pulse parameters were set. For box and line patterns, feedback remained on during lithography. Boxes were patterned in a serpentine manner with the tip moving in the long axis either along (parallel) or across (perpendicular) dimer rows\cite{Zyvex_HDL:2014}. Lithographic parameters and fit slopes, $m$, for sets of lines used to measure line width dependence on $V_{B}$ and $d_{e}$ are given in Table~\ref{tbl:bias} and Table~\ref{tbl:dose}, respectively. Line widths were determined by taking the full width at half max of the average line profiles across each patterned line.

Low-temperature experiments were performed using a separate cryogenic STM system. This system is home built and consists of a UHV sample preparation chamber with a base pressure $P$ \mbox{$< 1 \times 10^{-8}$ Pa} (\mbox{$7.5 \times 10^{-11}$ Torr}) and a cryogenic STM stage with a base temperature of \mbox{4.2 K}\cite{our_4K_system}. The sample used for these experiments consisted of a \mbox{1.5 $\mu$m} thick, high-purity intrinsic Si epilayer grown using molecular beam epitaxy on a highly P-doped Si(100) substrate with $\rho <$ \mbox{0.0015 $\Omega\cdot$cm}. The substrate was annealed to promote diffusion of dopants into the intrinsic Si layer. The sample surface and Cl-termination were prepared in the UHV chamber in a similar manner as in the other system mentioned earlier before transferring the sample to the cryogenic STM stage for low-temperature measurements including lithography and STM spectroscopy.

\begin{table}[b!]
  \caption{Lithographic patterning parameters including sample bias, $V_{B}$, and current setpoint, $I_{t}$, as well as linear fit slopes, $m$, for sets of lines patterned as a function of electron dose, $d_{e}$, in Fig.~\ref{sweeps}(d).}
  \label{tbl:dose}
  \begin{tabular}{llll}
    \hline
    data set                              & $V_{B}$ (V) & $I_{t}$ (nA) & $m$ (nm/(mC/cm)) \\
    \hline
    circles                               & $+$8.0        & 10           & 0.15(3)        \\
    squares                               & $+$8.0        & 5            & 0.16(2)        \\
    triangles                             & $+$9.0        & 5            & 0.13(2)        \\
    H-Si(100) (Ref.~\citenum{Lyding1996}) & $-$6.0        & 0.1          & 0.22           \\
    \hline
  \end{tabular}
\end{table}

\section{acknowledgement}
The authors thank C. J. K. Richardson (Laboratory for Physical Sciences) for providing the Si epilayer substrate used for low-temperature measurements.

\bibliographystyle{apsrev4-1}
\bibliography{Cl_Litho}

\end{document}